\documentstyle{europhys}

\def\etal{{\hbox{{\tenit\ et al.\/}\tenrm :\ }}}

\def\And{{\rm and\ }}

\def\stars{\bigskip\centerline{***}\medskip}

\newif\ifboo \boofalse

\def\Review#1{\boofalse{\it #1},}
\def\Name#1{{\sc #1},}
\def\Vol#1{\ifboo Vol. {\bf #1}\else{\bf #1}\fi}
\def\Year#1{\ifboo #1\else(#1)\fi}
\def\Book#1{\bootrue{\it #1},}
\def\Page#1{\ifboo {\rm p. #1}\else{\rm #1}\fi}

\input{psfig.sty}

\begin{document}

\euro{41}{4}{431-436}{1998}
\Date{15 February 1998}
\shorttitle{G. S. UHRIG \etal A MAGNETIC MODEL FOR THE INCOMMENSURATE ETC.}

\title{A magnetic model for the incommensurate {\it I} phase of 
spin-Peierls systems}

\author{G. S. Uhrig\inst{1}, F. Sch\"onfeld\inst{1} and 
J. P. Boucher\inst{2}}
\institute{
\inst{1}Institut f\"ur Theoretische Physik, Universit\"at zu K\"oln,
         D-50937 K\"oln, Germany.\\
\inst{2}Laboratoire de Spectrom\'etrie Physique, Universit\'e J. Fourier
 Grenoble I\\ BP 87, F-38402 Saint-Martin d'H\`eres cedex, France}

\rec{25 August 1997}{in final form 17 December 1997}

\pacs{
\Pacs{75}{40Gb}{Dynamic properties (dynamic susceptibility, spin waves,
 spin diffucsion, dynamic scaling etc.)}
\Pacs{75}{10Jm}{Quantized spin models}
\Pacs{75}{50Ee}{Antiferromagnetics}
      }
\maketitle

\begin{abstract}
A magnetic model is proposed for describing the incommensurate {\it I}
phase of spin-Peierls systems. Based on the harmonicity of the lattice
distortion, its main ingredient is that the distortion of the lattice
adjusts to the average magnetization such that the system is always 
gapful. The presence of dynamical incommensurabilities in the fluctuation
spectra is also predicted. Recent experimental results for CuGeO$_3$ 
obtained by NMR, ESR and light scattering absorption are well understood
within this model.
\end{abstract}

Since the discovery of the inorganic compound CuGeO$_3$ [1], much work is
devoted to the spin-Peierls transition, which is to be viewed as a
magneto-elastic distortion induced by quantum magnetic fluctuations [2], [3].
It is usually observed in $s = 1/2$ antiferromagnetic (AF) chains with
{\it isotropic} spin-spin interactions. For such uniform chains --- the
corresponding lattice structure defines the {\it U} phase of the system --- the
interplay between spins and phonons gives rise to a lattice distortion at
low temperature ($T \ll J$). Depending on the value of the applied magnetic
field $H$, this distortion results in a new lattice structure which remains
commensurate or becomes incommensurate. In low fields, the distortion
corresponds to a lattice dimerisation --- this structure defines the {\it D} 
phase. In large fields, the lattice incommensurability --- which defines the 
{\it I} phase --- increases with {\it H} [2]. 
Our purpose is to discuss for that {\it I} phase the
properties of a model Hamiltonian able to describe the features observed
experimentally. Comparisons with recent data obtained on CuGeO$_3$ are finally
presented.

As a result of the incommensurate lattice distortion, the magnetic
Hamiltonian in the {\it I} phase should display similar incommensurate
periodicities. For $H > H_c$, where $H_c$ is the critical field of the first
 order
transition between the {\it D} and {\it I} phases, we propose to describe the
 exchange coupling by the following Hamiltonian
\begin{equation}
        H = J \sum_i [1 - \delta_1 \cos(qr_i)] S_iS_{i+1}     
\end{equation}
where $\delta_1$ is the ``alternation" of the next-nearest exchange 
coupling $J$ and $q$
the wavevector characterizing the magnetic modulation. Higher
anharmonicities could easily be accounted for by adding terms 
$-\delta_n \cos(n q r_i)$ --- with $n$ odd integer --- in the square bracket of
 (1). For $q = \pi$, the
modulation is commensurate and the {\it D} phase is recovered. The case 
$\delta_1 = 0$ in (1) corresponds to the {\it U} phase. In the {\it I} phase, 
$q \neq \pi$ holds and we define $q = d + \pi$ where $d$ evaluates the lattice
 incommensurability, which is related to the average magnetization per site 
$\langle m\rangle = N^{-1}\sum_i m_i$ ($N$ is the system size).
In the $XY$ limit of (1), for $\langle m\rangle \neq 0$, the model is
 equivalent to tight-binding fermions with the  creation (annihilation) 
operators $c_i^\dagger\; (c_i)$ [6]-[8].  When an infinitesimal spin-lattice 
interaction $g$ is present, instabilities occur in  
the system [2]. The largest one develops at $q = 2k_{\rm F} = 
2\pi\langle m\rangle + \pi$, where the magnetic incommensurability 
$d = 2\pi\langle m\rangle$ is
created. Due to this instability, an energy gap 
$\Delta_+ + \Delta_-$ opens in reciprocal
space exactly at the two Fermi points as shown in fig. 1a). The states above
and below these gaps stay separated since they evolve {\it continuously} on
increasing $g$, i.e., no state can ``jump" across the gap. This picture remains
valid when interactions between fermions are taken into account --- i.e., in
the {\it isotropic} Heisenberg limit of (1). 
The corresponding chemical potential
$\mu$ is smaller than the Zeeman energy $h = g\mu_{\rm B}H$, as the
 interactions create an
internal field opposed to $H$. In the {\it I} phase, $\mu$ does not need to 
lie exactly in the middle of the energy gap (see fig. 1a)), i.e. $\Delta_+ \neq
\Delta_-$ in general.
Three different gaps have to be considered: $\Delta_+, \Delta_-$ and $\Delta_0$,
 which occur in
the energy spectra of the spin fluctuations $S_\perp(q,E)$ and 
$S_{||}(q,E)$ observed in
directions perpendicular and parallel to the applied field, respectively
(see figs. 1b,c).

Since the system is gapful, there exists an infrared cutoff and a
conventional perturbation treatment [8] offers a reliable approach. The
local magnetization $m_i = \langle c_i^\dagger c_i\rangle - 1/2$
  is evaluated as follows. In the $XY$
limit, a chain of tight-binding fermions with hopping $t_i$ and local energies
$\mu_i$ is considered. For each bond $i$, an effective two-site Hamiltonian 
$H_i$ and on-site and nearest-neighbor Green functions are defined via
\begin{equation}
H_i =\left(\begin{array}{cc}
\mu_i + L_i(E) & t_i \\
t_i & \mu_{i+1} + R_{i+1}(E)
\end{array}\right) \quad \mbox{and}\quad
\left(\begin{array}{cc}
G_{i;i} &  G_{i;i+1}\\
G_{i+1;i} & G_{i+1;i+1}
\end{array}\right) = (E-H_i)^{-1}
\end{equation}
The self-energy $L_i(E)$ ($R_{i+1}(E)$) takes the half-chain to the left (to the
right) of bond $i$ into account. The conditions 
$L_i = t_{i-1}^2/(E - \mu_{i-1} - L_{i-1})$ 
and $R_i = t_i^2/(E - \mu_{i+1} - R_{i+1})$ define recursively continued 
fraction
representations. The expectation values $\langle c_i^\dagger c_j\rangle$
 are computed  by the integral
$\langle c_i^\dagger c_j \rangle = (2\pi)^{-1} \int_{\Gamma_h}
 G_{i;j}(E)dE$, where the complex contour $\Gamma_h$ contains the
real interval $(- \infty, \mu]$.
 With $\mu$ within the gaps (see fig. 1a)), this integral
is easily evaluated. In the {\it isotropic} Heisenberg model, a Hartree-Fock (HF)
treatment is used. The fermionic chain is defined with 
$t_i = -J/2 - sJ\langle c_i^\dagger c_{i+1}\rangle$
 and $\mu_i = h - sJ(m_{i-1} + m_{i+1})$ [8]. The renormalization $s<1$ is
assumed to be independent of $d$ and $H$ for small fields ($h \ll J$).
 It is chosen as in ref. [8], such that in the {\it D} phase --- 
the $h,d = 0$ limit of our model ---
$\Delta_0 = \Delta_+ = \Delta_-$.
 The self-consistent HF calculation is done here with {\it full
spatial} dependence unlike the approach of Fujita and Machida (FM) [9] where
only a constant Fock term is considered. In the $XY$ limit, one obtains the
results shown in fig. 2a. The local magnetization $m_i$ has an alternating part
and a slowly varying one, which is spatially modulated with periodicity L/2
($L = 2\pi/d$). With interactions between fermions --- i.e. in the 
{\it isotropic}
Heisenberg limit of (1) --- these two important features are kept. However,
the $m_i$ become (alternately) negative --- see fig. 2b) --- and their amplitude is
much larger. For the same model (1), we performed also density-matrix
renormalisation group (DMRG) calculations on finite chains [11]. The results
confirm our renormalised HF approach very well. Note that such antiparallel
local magnetizations are generic to inhomogeneous AF chains [10].

From the fermionic dispersion sketched in fig.1a), a representation of the
spin fluctuation spectra in the {\it I} phase can be proposed. Based on arguments
similar to those used for uniform chains [7], [12], one is led for 
$S_{||}(q,E)$, and $S_\perp(q,E)$ to figs. 1b) and c), respectively.
 As in [12], one has to
distinguish between 
$S_{-+}(q,E) = \langle S^-(q)[E - ({\bf H} - E_0)]^{-1}S^+(q)\rangle$ and 
$S_{+-}(q,E) =
\langle S^+(q)[E - ({\bf H} - E_0)]^{-1}S^-(q)\rangle$ due to the broken spin
 rotation symmetry.
Accordingly, $S_\perp(q,E)$ is defined as $S_\perp(q,E) = 
[S_{-+}(q,E) + S_{+-}(q,E)]/2$. The
description for the spins is obtained through $S_i^+ = c_i^\dagger 
\exp[i\pi\sum_{l=-\infty}^{i-1}(1 -
c_l^\dagger c_l)]$. In mean-field treatment, this phase factor can 
be written as $\approx \exp[i\pi\sum_{l=-\infty}^{i-1}(1 - 
\langle c_l^\dagger c_l \rangle)]$. With $\langle c_l^\dagger c_l \rangle
\approx 1/2 + \langle m\rangle$ , it is seen to lead to
a wave vector shift $(\pi - d)/2$ [7]. Hence, for a given value of $d$, the 
lowest energy branch of $S_{||}(q,E)$ develops an incommensurate feature at 
$q = \pi - d$, where the energy gap $\Delta_0$ occurs also (see fig. 1b).
 In general, $\Delta_0 < \Delta_+ + \Delta_-$.
The same gap $\Delta_0$ appears at $q = 0$,
 while a Zeeman shift $h$ develops at $q =\pi$
(as in the {\it U} phase [12], it implies $\mu = h/2$). 
For $S_\perp(q,E)$ (fig. 1c), similar
spectra are obtained but shifted by $\pi$, with, however, the occurrence of the
gaps $\Delta_+$ and $\Delta_-$ at $q = d$. 
The Zeeman shift is found now at the center of  the
Brillouin zone ($q = 0$). In the inset of fig. 1b),c), the gap 
$\Delta_+ + \Delta_-$ is
calculated as a function of $\langle m\rangle$ for the values 
$J = 120$K, $\delta_1 = 0.12$
(corresponding to CuGeO$_3$ [3]) and $\delta_3 = - 0.07\delta_1$ (see below).
 Renormalized HF
(solid line) and DMRG (circles) results are in qualitative agreement. By
DMRG, it was also verified that $\Delta_+$ and $\Delta_-$ are indeed different,
 but of the same order of magnitude as in fig. 1c). The same approaches 
correctly predict
for the {\it D} phase [8] a triplet excited state
 giving a unique gap for $H = 0$:
$\Delta_0 = \Delta_+ = \Delta_- \sim 0.37 J \sim 44$K [13]. For $H \neq 0$,
 a Zeeman splitting occurs  and
three branches are observed as in CuGeO$_3$ [3].

The lattice incommensurability in the {\it I} phase of CuGeO$_3$ has been
 established
by X-ray measurements [4]. A small anharmonicity has also been observed [5].
However, the intensity ratio between the first and the third harmonic
super-lattice peaks is small ($I_3/I_1 \sim 1/200$ [5]), which let us expect
 only small anharmonicities in the exchange coupling ($\delta_3/\delta_1 =
 -\sqrt{I_3/I_1} \sim - 0.07$).
Frustrating next-nearest neighbor interactions $J_2$ ($> 0$) have also been
considered by DMRG. As shown in fig. 2b) for $\alpha = J_2 /J = 0.35$ [15] and
 for the same gap in the {\it D} phase ($E \sim 44$K) the results do not 
change very much.
For the local magnetization $m_i$ in the {\it I} phase, we refer to
 recent high field
nuclear magnetic resonance (NMR) measurements [16]. Above $H_c$ ($\sim 12.5$T
 in CuGeO$_3$) and below $T_{\rm sp}$ --- $T_{\rm sp}$($\sim 10$K 
for $H \sim 15 $T) is the critical temperature of the second order transition 
between the {\it U} and {\it I} phases --- a
distribution of the $m_i$ develops in the crystal. Below 4K, the NMR spectra
(fig. 4 in [16]) become practically $T$ independent. In other words, the
distribution of the local magnetization becomes quasi-static, and our $T =0$
calculation can be used. In [16], the NMR data were analyzed within the FM
soliton model [9]. In that model, the lattice distortion refers explicitly
to the sine-Gordon equation (which predicts too large 
$\delta_3/\delta_1$ ratios [5]) and the spin system is treated in the $XY$ 
limit except for a {\it constant} Fock renormalization of the coupling (for the 
{\it D} phase, the $XY$ limit predicts  only
a doublet state, not a triplet as observed in CuGeO$_3$ [3]). The FM approach
is therefore inadequate for CuGeO$_3$. As shown in fig. 2b),
 the {\it isotropic}
Heisenberg limit of (1) predicts a strongly negative part for $m_i$, which
seems to contradict the NMR data [16] where the distribution of the local
field is observed to be mainly positive and to have a much smaller amplitude
($\Delta S^z=S^z_{\rm max} - S^z_{\rm min}\sim 0.065$). 
This important reduction of the experimental value of $\Delta S^z$ can be
explained by the zero-point motion of the so-called phasons [17].
Such gapless excitations are linked to the possibility for the incommensurate 
distortion to slide along the chain without energy cost. This non-adiabatic
effect induces ``oscillations" of the magnetic pattern, which reduce the 
alternating component of $m_i$ and result in an averaging over adjacent
sites as $m_i^{\rm eff} = (1-2\gamma)m_i + \gamma(m_{i+1}+m_{i-1})$.
In the present work, $\gamma$ is used as a fit parameter, though a quantitative
evaluation of the averaging, which is to be presented elsewhere, gives correct
orders of magnitude.  From the diamond
data in fig. 2b) ($\alpha = 0$, {\it i.e.} no frustration) one obtains for 
$\gamma  = 0.20$ the
effective local magnetization depicted in fig. 2c).
 From the square data ($\alpha = 0.35$) of fig. 2b) almost identical results
 (not shown) are obtained for $\gamma =0.19$. The averaging restores a mainly
 positive distribution, which is
remarkably similar to that obtained in the $XY$ limit (compare figs. 2a) and
c)). Following the same procedure as in [16], we obtain from fig. 2c) the NMR
lineshapes displayed in fig. 3. A reasonable agreement is achieved for 
$\delta_1 = 0.12$, a result consistent with the value determined previously
 for CuGeO$_3$
[14]. For the intrinsic damping of the NMR line, we took $\sigma = 0.035$T.
 This damping partly results from the experimental procedure since, in [16], the
NMR signal was recorded as a function of $H$. During the field sweep, the
incommensurability $d$ is varied, changing continuously the distribution of
the $m_i$. This could explain the apparent ``training" observed at low and high
fields.

Concerning the spin dynamics in the {\it I} phase, we refer to inelastic Raman
scattering (IRS) [18], [19] and electron spin resonance (ESR) [20] measurements
performed in CuGeO$_3$ well above $H_c$. Concerning IRS, for any field values a
striking result is obtained: a peak at $E \sim 230$ cm$^{-1}$ is observed in
 all the
three phases [18], [19]. Such a peak occurs when the density of states of
elementary excitations becomes large. In the {\it D} phase, it corresponds to
(twice) the maximum of the triplet $S = 1$ elementary excitations (fig.1a) in
[18]), and in the {\it U} phase, to (twice) the maximum ($E_{\rm max}$) of the
 low energy
spinon branch (fig.1b) in [18]). The peak observed in the {\it I} phase
 (fig. 2 in
[19]) is noticeably similar to the peak in the {\it U} phase. This strongly
supports the representations proposed in figs. 1b) and c) where both the {\it U}
 and
I phases are represented: at high energy, the excitation branches behave
similarly, with the same $E_{\rm max}$ and therefore the same density of 
states. In an ESR measurement, the $q = 0$ mode of $S_\perp(q,E)$ is directly 
probed. In high
fields, a remarkable behavior of the ESR signal has been reported [20]: no
change on its position and lineshape occurs when the {\it U}-{\it I} transition
 line is crossed (fig. 1 in [20]). This result supports well our
 conjecture presented
in fig.1c) since the $q = 0$ mode undergoes the same Zeeman shift $h$ 
in the two
phases. At $T_{\rm sp}$, the only dynamical changes
 occur in the low energy part of
the spectra, with the opening of  the energy gaps $\Delta_+$ and $\Delta_-$.

In conclusion, a magnetic Hamiltonian is proposed to describe the properties
of the {\it I} phase of a spin-Peierls system [21]. The lattice distortion is
considered to result mainly in a modulation of the exchange alternation.
Both the local magnetization distribution and the dynamical properties known
at present for the {\it I} phase of CuGeO$_3$ are well explained by this model.
 It
predicts the adiabatic {\it I} phase to be characterized at low energy by two
important features: the presence of a dynamical incommensurability (as in
the {\it U} phase) and the opening of energy gaps (as in the {\it D} phase).
 Varying the
external field tends to reduce either $\Delta_+$ or $\Delta_-$.
 The whole spin-lattice
system, however, responds by changing its global modulation in order to
minimize its total energy so that eventually none of the gaps closes.

\stars
We would like to thank Y. {\sc Fagot-Revurat}, H. J. {\sc Schulz} and 
{\sc Th. Nattermann} for helpful
discussions. One of us (GSU) acknowledges the hospitality of Laboratoire de
Physique des Solides, Universit\'e Paris-Sud, Orsay, where this work was
initiated, and the financial support of the Deutsche Forschungsgemeinschaft
(individual grant and SFB 341).

\newpage

\begin{figure}
\centerline{\psfig{file=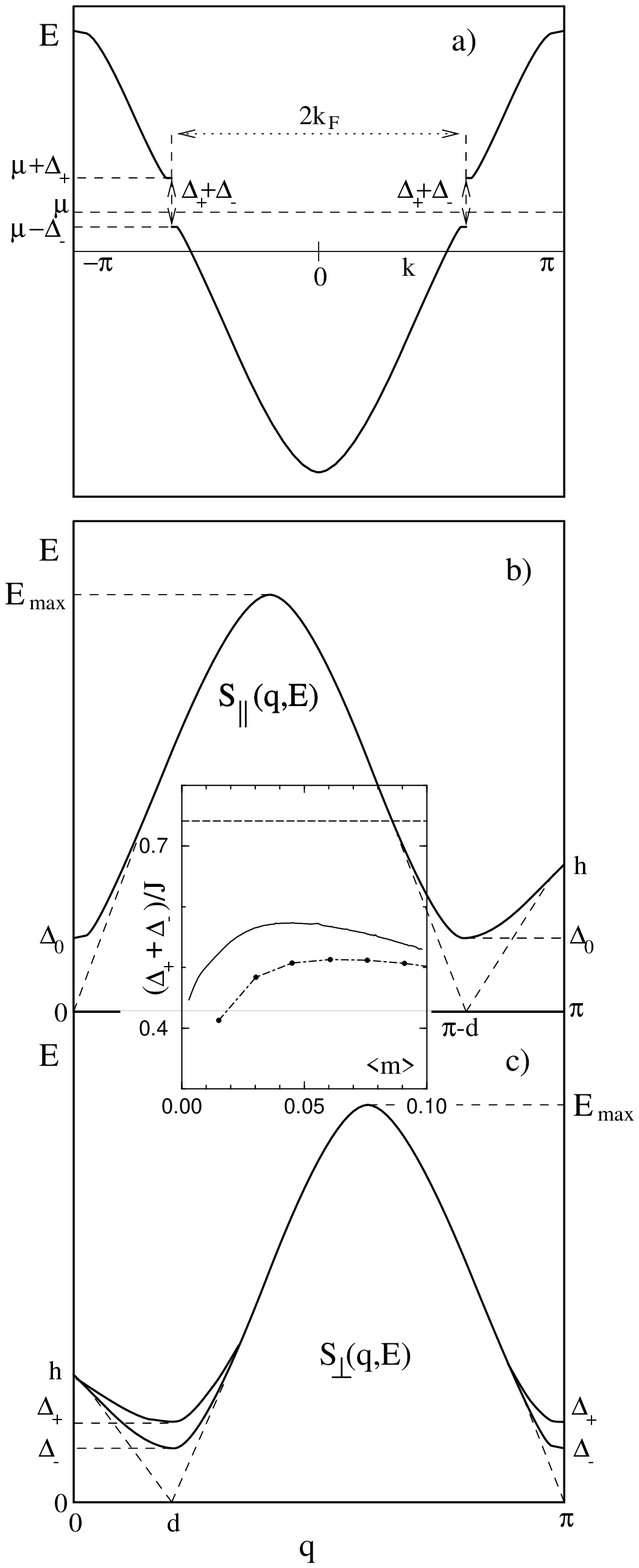,width=14cm}}
\caption{
\label{fig1}
a) Fermionic dispersion for Hamiltonian (1). The Fermi energy is
denoted by $\mu$; b) and c) dispersions of the spin excitations in 
$S_{||}(q,E)$ and $S_\perp(q,E)$ for the {\it I} and {\it U} phases (solid and
 dashed  lines, respectively). For $h \ll J, E_{\rm Max}/J \sim (\pi/2)$.
 Inset: dependence of  the energy gap $\Delta_+ + \Delta_-$ as a
function of the incommensurability $d$ (solid line), for 
$\delta_1 = 0.12$ [14], $\delta_3 = -0.07\delta_1$ and $\alpha= 0$ (see text).
 The dots depict DMRG results for a 66 site
chain. The dashed line is twice the gap value ($0.37J$) determined for
 the {\it D} phase [14].
}
\end{figure}

\begin{figure}
\centerline{\psfig{file=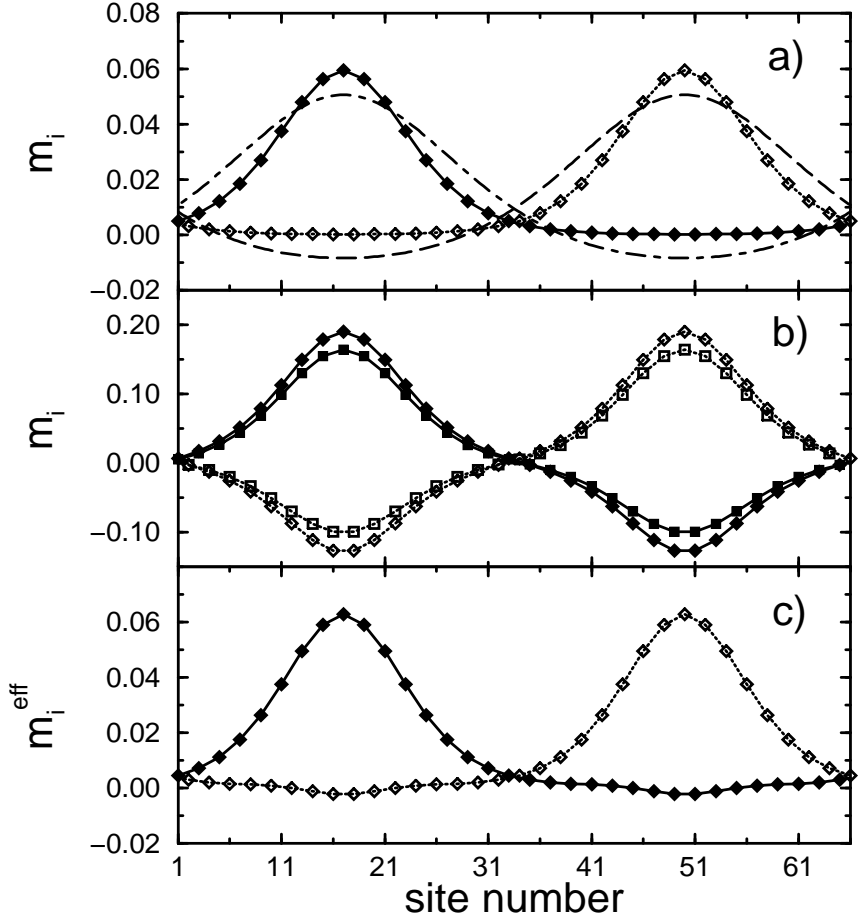,width=14cm}}
\caption{
\label{fig2}
a) Local magnetization versus site number in the $XY$ limit of
Hamiltonian (1) for $L (= 2\pi/d) \sim 66$ (as in [16]), 
$\delta_1 = 0.12, \delta_3 = -0.07\delta_1$
and $\alpha = 0$; b) diamonds: as previously but in the Heisenberg limit of (1)
(DMRG provides almost identical results); squares: DMRG results (same gap
but $\alpha = 0.35, \delta_1 = 0.033, \delta_3 = -0.07\delta_1$ for a 66 site
 chains ; c) the effective local magnetization in the isotropic Heisenberg
 limit for $\gamma = 0.20$ (see text). In a), the dashed and dot-dashed lines
 correspond to the FM model [9] calculated for the value $k=0.9$ 
 ($k$ is the modulus of the elliptic
Jacobi function; the offset occuring in the FM description is determined by
$\langle m\rangle = 1/L$ and $L=66$.
 Full symbols: odd sites; empty symbols: even sites;
curves through symbols are guides to the eye.}
\end{figure}

\begin{figure}
\centerline{\psfig{file=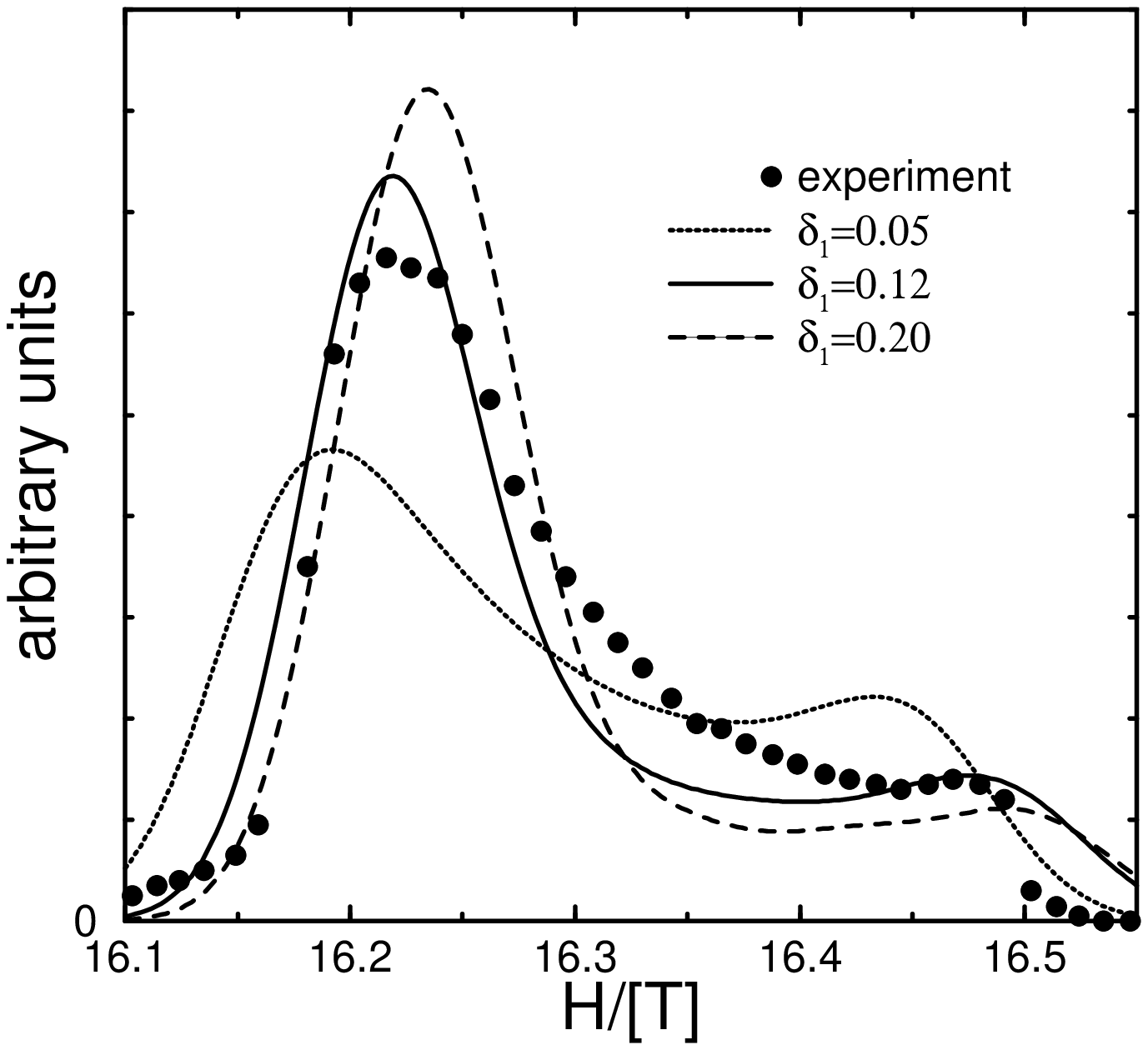,width=14cm}}
\caption{
\label{fig3}
NMR lineshapes evaluated from the distribution of the local
magnetization displayed on fig. 2c), for different values of $\delta_1$ and
$\delta_3 = -0.07 \delta_1$. The internal NMR damping is 
$\sigma = 0.035$T (see text).}
\end{figure}


\begin{thebibliography}{10}
\bibitem{13} 
This value agrees well with the experiments in CuGeO$_3$ [3]: calculated
for a purely 1D system, it should be compared not to the lowest gap 
($\sim 23$ K) but to an ``average gap", the averaging being made over the 
dispersions transverse to the chains [14].
\bibitem{21} In the {\it I} phase, we do not expect important changes for the 
local
magnetization from the interchain couplings. They will, however, reduce the
gap values as they do in the {\it D} phase [14].
\end{thebibliography}
\end{document}